\documentclass[12pt,a4paper]{article}
\usepackage{amssymb}
\usepackage{amsmath}
\usepackage{latexsym}
\usepackage{cite}
\usepackage{physics}
\usepackage{float}
\usepackage{graphicx}
\begin{document}

\title{\vspace{-1cm}\bf Peculiarities of quantum field theory\\ in the presence of a wormhole}

\author{
Mikhail~N.~Smolyakov
\\
{\small{\em Skobeltsyn Institute of Nuclear Physics, Lomonosov Moscow
State University,
}}\\
{\small{\em Moscow 119991, Russia}}}

\date{}
\maketitle

\begin{abstract}
In this paper, quantum theory of a real massive scalar field in the background of a traversable wormhole is examined. The wormhole is supposed to connect two different universes; as a particular example the simplest Ellis wormhole is considered. It is shown that the resulting theory possesses a doubling of quantum states parametrized by the same value of asymptotic momentum; two different sets of such degenerate states are localized in different universes. This is a consequence of the topological structure of the wormhole spacetime, which is the same as the topological structure of the Schwarzschild spacetime providing a similar degeneracy of quantum states.
\end{abstract}

\section{Introduction}
The question of quantization of fields in a curved background is one of the most discussed in modern theoretical physics. The amount of scientific literature on this topic is very large; the key starting points can be found, for example, in the classical textbook \cite{BD}. After the well-known papers \cite{Boulware:1974dm,HH}, the main attention is attracted to black holes. However, in principle there may exist other quite exotic objects which also deserve examination.

In the present paper one such object is considered --- a traversable wormhole \cite{Ellis:1973yv,Bronnikov:1973fh,Morris:1988cz}. An important point is that such a wormhole is supposed to connect two different universes, not two different areas of our Universe. As a particular example for which all calculations will be performed, the Ellis wormhole \cite{Ellis:1973yv}, which is the simplest traversable wormhole, is chosen. We will not be interested in how such a wormhole could be created, but just consider the corresponding metric as a background.

An important property of such wormholes is that topologically the corresponding spacetime is $R^{2}\times S^{2}$ (the topology of Minkowski spacetime is $R^{4}$). Recall that the Schwarzschild spacetime has exactly the same topology. In papers \cite{Egorov:2022hgg,Smolyakov:2023pml} it was shown that there exists an additional degeneracy of quantum states in comparison with the case of Minkowski spacetime, which is the consequence of the Schwarzschild spacetime topology. It was noted in \cite{Egorov:2022hgg,Smolyakov:2023pml} that analogous degeneracy is expected for the cases of wormholes connecting two different universes. The analysis presented below demonstrates that it is indeed so --- the complete set of physical states of the resulting quantum theory is split into two sets of states, each of which lives mainly in its own universe. Since the Ellis wormhole is very simple, has no horizon and long-range gravitational potential, and possesses an additional symmetry in comparison with the Schwarzschild black hole (namely, the potential in the corresponding radial equation is even in the radial coordinate), the analysis and results seem to be even more demonstrative than those for the Schwarzschild black hole and can be considered as a simpler realization of the effect described in \cite{Egorov:2022hgg,Smolyakov:2023pml}. A real massive scalar field will be used to examine the features of quantum field theory on such a background. A possible backreaction of the scalar field on the background metric is not taken into account.

The paper is organized as follows. In Section~\ref{sectsetup}, the metric and the action of the scalar field are introduced. In Section~\ref{sectnorm}, the properties of radial solutions are discussed in detail. In Section~\ref{sectscatstates}, the scattering states are composed using the radial solutions obtained in Section~\ref{sectnorm}. In Section~\ref{sectnewstates}, the physical states that are combinations of the scattering states are defined and their properties are examined. In Section~\ref{sectqft}, a consistent quantum field theory for the scalar field is built. In Section~\ref{sectconclusion}, the results obtained in the present paper are discussed. The Appendix contains auxiliary material.

\section{Setup}\label{sectsetup}
The metric of the Ellis wormhole has the form \cite{Ellis:1973yv}
\begin{equation}\label{metric_WH}
ds^2=dt^2-dr^2-(r^2+b^{2})\left(d\theta^2+\sin^2\theta\,d\varphi^2\right),
\end{equation}
where $r\in (-\infty,\infty)$ and $b^{2}>0$. Let us consider a real massive scalar field $\phi(t,r,\theta,\varphi)$ in a curved background described by metric \eqref{metric_WH}:
\begin{equation}\label{scalact}
S=\int\sqrt{-g}\left(\frac{1}{2}\,g^{\mu\nu}\partial_{\mu}\phi\,\partial_{\nu}\phi-\frac{M^{2}}{2}\phi^{2}\right)d^{4}x.
\end{equation}
Since the metric is static, the equation of motion following from action \eqref{scalact} takes the form
\begin{equation}\label{scalareqm}
\sqrt{-g}\,g^{00}\ddot\phi+\partial_{i}\left(\sqrt{-g}\,g^{ij}\partial_{j}\phi\right)+M^{2}\sqrt{-g}\,\phi=0,
\end{equation}
where $\dot\phi=\partial_{0}\phi$.

\section{Radial solutions}\label{sectnorm}
The metric \eqref{metric_WH} is spherically symmetric, so as a starting step it is natural to expand the scalar field $\phi(t,r,\theta,\varphi)$ in solutions of the form
\begin{equation}\label{philm}
e^{\pm iE t}\phi_{lm}^{}(E,r,\theta,\varphi)=e^{\pm iE t}Y_{lm}(\theta,\varphi)f_{l}(k,r),
\end{equation}
where
\begin{equation}\label{Ylm}
Y_{lm}(\theta,\varphi)=\sqrt{\frac{2l+1}{4\pi}}\sqrt{\frac{(l-|m|)!}{(l+|m|)!}}\,P_{l}^{|m|}\left(\cos\theta\right)e^{im\varphi},\quad l=0,1,2, ... ,\quad m=0,\pm 1, \pm 2, ...
\end{equation}
are spherical harmonics in the convention of \cite{Korn-Korn}. Without loss of generality, we assume that $E\ge 0$. The parameter $k$ is defined as $k=\sqrt{E^{2}-M^{2}}$. Substituting representation \eqref{Ylm} into Eq.~\eqref{scalareqm}, one gets the radial equation
\begin{equation}\label{eqscalarrad}
\left(E^{2}-M^{2}\right)f_{l}(k,r)+\frac{1}{r^{2}+b^{2}}\frac{d}{dr}\left(\left(r^{2}+b^{2}\right)\frac{df_{l}(k,r)}{dr}\right)
-\frac{l(l+1)}{r^{2}+b^{2}}f_{l}(k,r)=0
\end{equation}
for the functions $f_{l}(k,r)$. Since the coefficients in Eq.~\eqref{eqscalarrad} are real, the functions $f_{l}(k,r)$ can be chosen to be real as well.

\begin{figure}[ht]
\centering
\includegraphics[width=0.9\linewidth]{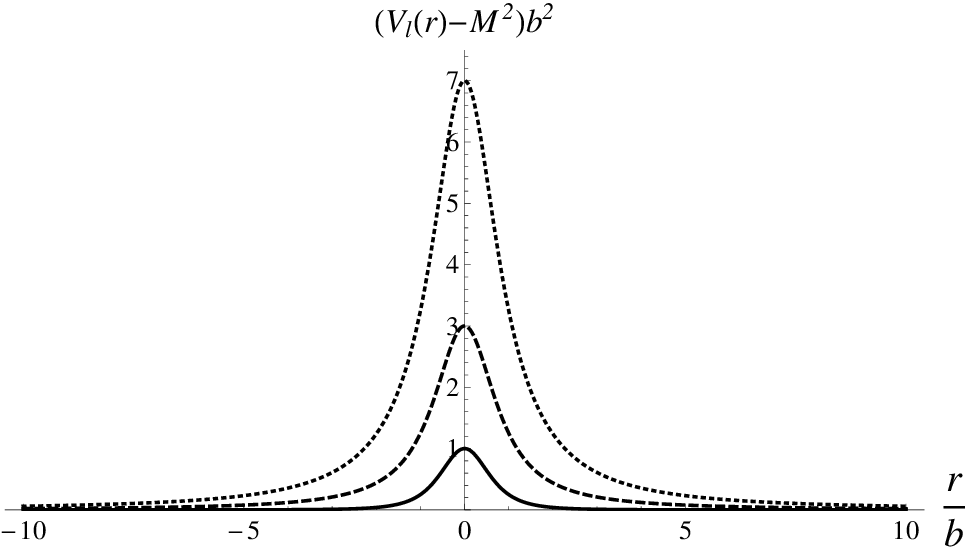}
\caption{$V_{l}(r)$ for $l=0$ (solid line), $l=1$ (dashed line), and $l=2$ (dotted line).}\label{fig}
\end{figure}

Let us introduce the new function
\begin{equation}\label{substdimens}
\psi_{l}(k,r)=\sqrt{r^{2}+b^{2}}f_{l}(k,r).
\end{equation}
With \eqref{substdimens}, Eq.~\eqref{eqscalarrad} reduces to a one-dimensional Schr\"{o}dinger equation,
\begin{equation}\label{eqSchr}
-\frac{d^{2}\psi_{l}(k,r)}{dr^{2}}+V_{l}(r)\psi_{l}(k,r)=E^{2}\psi_{l}(k,r),
\end{equation}
with the potential \cite{Clement:1982ej,Kar:1994ty}
\begin{equation}\label{VSchr1}
V_{l}(r)=M^{2}+\frac{b^{2}}{\left(r^{2}+b^{2}\right)^{2}}+\frac{l(l+1)}{r^{2}+b^{2}}
\end{equation}
such that $V_{l}(r)\to M^{2}$ for $r\to\pm\infty$. In Fig.~\ref{fig}, examples of $V_{l}(z)$ for $l=0$, $l=1$, and $l=2$ are presented.

Properties of eigenfunctions of the problem with potential \eqref{VSchr1} are quite clear. First, nonzero solutions exist for $E>M$ (which means that $k$ is real) --- for any $k>0$ and fixed $l$ there exist two linearly independent solutions which are bounded for $r\to\pm\infty$. As was noted above, they can be chosen to be real. Second, potential \eqref{VSchr1} is symmetric with respect to $r\leftrightarrow -r$. Thus, the two linearly independent solutions mentioned above can be chosen to be symmetric and antisymmetric in $r$:
\begin{align}\label{radsols}
&\psi_{l,s}(k,-r)=\psi_{l,s}(k,r),\\\label{radsola}
&\psi_{l,a}(k,-r)=-\psi_{l,a}(k,r).
\end{align}
The normalization and orthogonality conditions for these eigenfunctions take the form
\begin{align}\label{normq}
&\int\limits_{-\infty}^{\infty}\psi_{l,s}(k,r)\psi_{l,s}(k',r)dr=\delta(k-k'),\\
&\int\limits_{-\infty}^{\infty}\psi_{l,a}(k,r)\psi_{l,a}(k',r)dr=\delta(k-k'),\\\label{normq3}
&\int\limits_{-\infty}^{\infty}\psi_{l,s}(k,r)\psi_{l,a}(k',r)dr=0,
\end{align}
whereas the completeness relation has the form\footnote{Since Eq.~\eqref{eqSchr} with potential \eqref{VSchr1} for a fixed $l$ represents an eigenvalue problem of the Hermitian operator, the corresponding eigenfunctions form a complete set \cite{Korn-Korn}.}
\begin{equation}\label{compl}
\int\limits_{0}^{\infty}\Bigl(\psi_{l,s}(k,r)\psi_{l,s}(k,r')+\psi_{l,a}(k,r)\psi_{l,a}(k,r')\Bigl)dk=\delta(r-r').
\end{equation}

Let us examine the asymptotic behavior of the solutions $\psi_{l,s}(k,r)$ and $\psi_{l,a}(k,r)$. For $r\to\pm\infty$, Eq.~\eqref{eqSchr} with potential \eqref{VSchr1} reduces to the equation
\begin{equation}\label{eqasympinfty}
-\frac{d^{2}\psi_{l}(k,r)}{dr^{2}}+M^{2}\psi_{l}(k,r)=E^{2}\psi_{l}(k,r).
\end{equation}
Thus, the solutions $\psi_{l,s}(k,r)$ and $\psi_{l,a}(k,r)$ for $r\to\pm\infty$ can be represented as
\begin{align}\label{solz1}
\psi_{l,s}(k,r)&\approx C_{l,s}(k)\sin\left(k|r|+\kappa_{l,s}(k)\right),\\\label{solz2}
\psi_{l,a}(k,r)&\approx\textrm{sign}(r)C_{l,a}(k)\sin\left(k|r|+\kappa_{l,a}(k)\right).
\end{align}
Here the normalization constants $C_{l,s}(k)$ and $C_{l,a}(k)$ are supposed to be positive, and $\kappa_{l,s}(k)$ and $\kappa_{l,a}(k)$ are some phases.\footnote{Contrary to the case of the Schwarzschild black hole \cite{Egorov:2022hgg,Smolyakov:2023pml}, asymptotics \eqref{solz1} and \eqref{solz2} do not contain additional phase shifts $\sim\ln(kr)$ because there is no term $\sim\frac{1}{r}$ in potential \eqref{VSchr1}. Note that the phases $\kappa_{l,a}(k)$ (in our notations) were estimated in \cite{Clement:1982ej} (not the phases $\kappa_{l,s}(k)$, because in paper \cite{Clement:1982ej} an extra restriction on the scalar field was imposed which led to discarding radial solutions that are even in $r$ (which are $f_{l,s}(k,R)$ in our notations)).}

Let us calculate the normalization constants $C_{l,s}(k)$ and $C_{l,a}(k)$. It is well known that normalization constants for such eigenfunctions can be calculated using their explicit form only in the asymptotic regions; see the trick used in \S21 of \cite{LL-QM} for calculating normalization constants in a similar case. Let us start with calculating $C_{l,s}(k)$ and rewrite the integral in \eqref{normq} as
\begin{align}\nonumber
&\int\limits_{-\infty}^{\infty}\psi_{l,s}(k,r)\psi_{l,s}(k',r)dr\approx C_{l,s}(k)C_{l,s}(k')\int\limits_{-\infty}^{-L}
\sin\left(-kr+\kappa_{l,s}(k)\right)\sin\left(-k'r+\kappa_{l,s}(k')\right)dr\\\nonumber
&+C_{l,s}(k)C_{l,s}(k')\int\limits_{L}^{\infty}
\sin\left(kr+\kappa_{l,s}(k)\right)\sin\left(k'r+\kappa_{l,s}(k')\right)dr
+\int\limits_{-L}^{L}\psi_{l,s}(k,r)\psi_{l,s}(k',r)dr\\\label{normderiv1}
&=2\,C_{l,s}(k)C_{l,s}(k')\int\limits_{L}^{\infty}
\sin\left(kr+\kappa_{l,s}(k)\right)\sin\left(k'r+\kappa_{l,s}(k')\right)dr
+\int\limits_{-L}^{L}\psi_{l,s}(k,r)\psi_{l,s}(k',r)dr,
\end{align}
where $L$ is such that for $|r|>L$ asymptotic solution \eqref{solz1} can be used with a good accuracy.

The second integral in the last line of formula \eqref{normderiv1} is finite, and its contribution to the overall infinite value of the normalization integral is irrelevant. Thus, we can replace this finite integral in the last line of \eqref{normderiv1} by a different finite integral, i.e., by
\begin{equation}\label{finiteint}
\int\limits_{-L}^{L}\psi_{l,s}(k,r)\psi_{l,s}(k',r)dr\to 2\,C_{l,s}(k)C_{l,s}(k')\int\limits_{0}^{L}
\sin\left(kr+\kappa_{l,s}(k)\right)\sin\left(k'r+\kappa_{l,s}(k')\right)dr.
\end{equation}
As a result, for the normalization integral \eqref{normderiv1} we obtain
\begin{equation}
\int\limits_{-\infty}^{\infty}\psi_{l,s}(k,r)\psi_{l,s}(k',r)dr\approx 2\,C_{l,s}(k)C_{l,s}(k')\int\limits_{0}^{\infty}
\sin\left(kr+\kappa_{l,s}(k)\right)\sin\left(k'r+\kappa_{l,s}(k')\right)dr.
\end{equation}
For $k\to k'$, in the leading order the latter integral can be rewritten as
\begin{align}\nonumber
&2\,C_{l,s}^{2}(k)\int\limits_{0}^{\infty}
\sin\left(kr+\kappa_{l,s}(k)\right)\sin\left(k'r+\kappa_{l,s}(k)\right)dr\\\label{normderiv3}
&\approx-\frac{C_{l,s}^{2}(k)}{2}\int\limits_{0}^{\infty}
\Biggl(e^{i\left(2kr+2\kappa_{l,s}(k)\right)}
+e^{-i\left(2kr+2\kappa_{l,s}(k)\right)}-e^{i(k-k')r}-e^{i(k'-k)r}\Biggr)dr.
\end{align}
The first two terms in the brackets in the last line of formula \eqref{normderiv3} are purely oscillating for $k=k'$, and thus they are irrelevant for the calculation of the normalization constants $C_{l,s}(k)$. As for the relevant terms, they can be rewritten as
\begin{align}\nonumber
&\frac{C_{l,s}^{2}(k)}{2}\int\limits_{0}^{\infty}\Biggl(e^{i(k-k')r}+e^{i(k'-k)r}\Biggr)dr
=\frac{C_{l,s}^{2}(k)}{2}\int\limits_{0}^{\infty}e^{i(k-k')r}dr+\frac{C_{l,s}^{2}(k)}{2}\int\limits_{-\infty}^{0}e^{i(k-k')r}dr\\\label{normderiv4}
&=\frac{C_{l,s}^{2}(k)}{2}\int\limits_{-\infty}^{\infty}e^{i(k-k')r}dr=C_{l,s}^{2}(k)\,\pi\delta(k-k').
\end{align}
Taking into account \eqref{normq}, one obtains
\begin{equation}\label{Clsk}
C_{l,s}(k)=\frac{1}{\sqrt{\pi}}.
\end{equation}
Using \eqref{solz2}, a fully analogous procedure can be performed for $\psi_{l,a}(k,r)$, resulting in the same value of normalization constants
\begin{equation}\label{Clak}
C_{l,a}(k)=\frac{1}{\sqrt{\pi}}.
\end{equation}

\section{Scattering states}\label{sectscatstates}
At this point, it turns out to be more convenient to pass to the isotropic coordinates by means of the transformation \cite{Clement:1982ej}
\begin{equation}
r=R-\frac{b^{2}}{4R},\qquad R\in \left(-\infty,-\frac{b}{2}\right)\cup\left[\frac{b}{2},\infty\right).
\end{equation}
In these coordinates, the metric takes the form
\begin{equation}\label{metric_Sch_isotropic}
ds^2=dt^2-\left(1+\frac{b^{2}}{4R^{2}}\right)^{2}\left(dR^2+R^{2}(d\theta^2+\sin^2\theta\,d\varphi^2)\right),
\end{equation}
whereas Eq.~\eqref{eqscalarrad} takes the form
\begin{equation}\label{eqscalarrad-isotropic}
\left(E^{2}-M^{2}\right)f_{l}(k,R)+\frac{16R^{4}}{\left(4R^{2}+b^{2}\right)^{3}}\frac{d}{dR}\left(\left(4R^{2}+b^{2}\right)\frac{df_{l}(k,R)}{dR}\right)
-\frac{16R^{2}l(l+1)}{\left(4R^{2}+b^{2}\right)^{2}}f_{l}(k,R)=0.
\end{equation}
Here the first universe is the one with $R>\frac{b}{2}$, whereas the second universe is the one with $R<-\frac{b}{2}$.

Orthogonality conditions \eqref{normq}--\eqref{normq3} now look like
\begin{align}\label{normqК1}
&\int\limits_{-\infty}^{-\frac{b}{2}}f_{l,s}(k,R)f_{l,s}(k',R)\frac{\left(4R^{2}+b^{2}\right)^{3}}{64R^{4}}\,dR
+\int\limits_{\frac{b}{2}}^{\infty}f_{l,s}(k,R)f_{l,s}(k',R)\frac{\left(4R^{2}+b^{2}\right)^{3}}{64R^{4}}\,dR=\delta(k-k'),\\
&\int\limits_{-\infty}^{-\frac{b}{2}}f_{l,a}(k,R)f_{l,a}(k',R)\frac{\left(4R^{2}+b^{2}\right)^{3}}{64R^{4}}\,dR
+\int\limits_{\frac{b}{2}}^{\infty}f_{l,a}(k,R)f_{l,a}(k',R)\frac{\left(4R^{2}+b^{2}\right)^{3}}{64R^{4}}\,dR=\delta(k-k'),\\\label{normqК3}
&\int\limits_{-\infty}^{-\frac{b}{2}}f_{l,s}(k,R)f_{l,a}(k',R)\frac{\left(4R^{2}+b^{2}\right)^{3}}{64R^{4}}\,dR
+\int\limits_{\frac{b}{2}}^{\infty}f_{l,s}(k,R)f_{l,a}(k',R)\frac{\left(4R^{2}+b^{2}\right)^{3}}{64R^{4}}\,dR=0,
\end{align}
and completeness relation \eqref{compl} takes the form
\begin{equation}\label{comp2}
\int\limits_{0}^{\infty}\Bigl(f_{l,s}(k,R)f_{l,s}(k,R')+f_{l,a}(k,R)f_{l,a}(k,R')\Bigl)dk=\frac{64R^{4}}{\left(4R^{2}+b^{2}\right)^{3}}\delta(R-R').
\end{equation}
The asymptotics \eqref{solz1} and \eqref{solz2} with \eqref{Clsk}, \eqref{Clak}, and \eqref{substdimens} can be rewritten as
\begin{equation}\label{sol-large-r}
f_{l,s}(k,R)\approx\frac{1}{\sqrt{\pi}\,|R|}\sin\left(k|R|-\frac{\pi l}{2}+\delta_{l,s}(k)\right),
\end{equation}
\begin{equation}\label{sol-large-r2}
f_{l,a}(k,R)\approx\frac{\textrm{sign}(R)}{\sqrt{\pi}\,|R|}\sin\left(k|R|-\frac{\pi l}{2}+\delta_{l,a}(k)\right),
\end{equation}
where the phase shifts $\delta_{l,s}(k)$, $\delta_{l,a}(k)$ are expressed through the phase shifts $\kappa_{l,s}(k)$, $\kappa_{l,a}(k)$ as $\delta_{l,s}(k)=\kappa_{l,s}(k)+\frac{\pi l}{2}$ and $\delta_{l,a}(k)=\kappa_{l,a}(k)+\frac{\pi l}{2}$.

Let us define the combinations
\begin{align}\label{scatstatesdec00}
\phi_{s}(k,R,\theta)&=\frac{1}{4\pi k}\sum\limits_{l=0}^{\infty}(2l+1)e^{i\left(\frac{\pi l}{2}+\delta_{l,s}(k)\right)}P_{l}\left(\cos\theta\right)f_{l,s}\left(k,R\right),\\\label{scatstatesdec00asym}
\phi_{a}(k,R,\theta)&=
\frac{1}{4\pi k}\sum\limits_{l=0}^{\infty}(2l+1)e^{i\left(\frac{\pi l}{2}+\delta_{l,a}(k)\right)}P_{l}\left(\cos\theta\right)f_{l,a}\left(k,R\right),
\end{align}
where $P_{l}(...)$ are the Legendre polynomials. In fact, these states are constructed exactly in the same way as the standard scattering states in quantum mechanics (see, for example, \cite{LL-QM}). One can easily show that for large $|R|$
\begin{align}\label{scatstatesrtheta}
\phi_{s}(k,R,\theta)&\approx\frac{1}{\sqrt{2}(2\pi)^{\frac{3}{2}}}\left(e^{ik|R|\cos\theta}+A_{s}\left(k,\theta\right)\frac{e^{ik|R|}}{|R|}\right),\\\label{scatstatesrtheta2}
\phi_{a}(k,R,\theta)&\approx\frac{\textrm{sign}(R)}{\sqrt{2}(2\pi)^{\frac{3}{2}}}\left(e^{ik|R|\cos\theta}+A_{a}\left(k,\theta\right)\frac{e^{ik|R|}}{|R|}\right),
\end{align}
where
\begin{equation}\label{scatamplrtheta}
A_{p}\left(k,\theta\right)=\frac{1}{2ik}\sum\limits_{l=0}^{\infty}(2l+1)P_{l}\left(\cos\theta\right)\left(
e^{i2\delta_{l,p}(k)}-1\right),\qquad p=s,a
\end{equation}
have the same form as the standard scattering amplitudes \cite{LL-QM}.

For $R\to\infty$ the first terms in the brackets of formulas \eqref{scatstatesrtheta} and \eqref{scatstatesrtheta2} describe plane waves traveling along the axis $z=R\cos\theta$ (analogously for $R\to-\infty$). In the standard quantum mechanical case, generalization to an arbitrary direction along which a plane wave propagates is performed by passing to Cartesian coordinates and using the relation $\cos\theta=\frac{\vec k\vec r}{kr}$ \cite{LL-QM}. Here the situation is more involved, because we cannot choose one and the same Cartesian-like coordinates (i.e., coordinates that tend to Cartesian ones for $|R|\to\infty$) for $R>\frac{b}{2}$ and $R<-\frac{b}{2}$. To get around this problem, let us define the coordinates
\begin{equation}
\vec X=\left[\begin{array}{ll}
\vec x=(R\sin\theta\cos\varphi, R\sin\theta\sin\varphi, R\cos\theta),\qquad\quad\,\,\,\, R>\frac{b}{2},\\
\vec y=(|R|\sin\theta\cos\varphi, |R|\sin\theta\sin\varphi, |R|\cos\theta),\qquad R<-\frac{b}{2}.
\end{array}
\right.
\end{equation}
Coordinates $\vec x$ and $\vec y$ tend to Cartesian coordinates for $R\to\infty$ and $R\to-\infty$ respectively. In these coordinates, the equation for the spatial part of the scalar field $\phi(k,\vec X)$ takes the form
\begin{equation}\label{eqscalarrad-isotropic-Cart}
\left(1+\frac{b^{2}}{4\vec X^{2}}\right)^{3}\left(E^{2}-M^{2}\right)\phi(k,\vec X)-\eta^{ij}
\frac{\partial}{\partial X^{i}}\left(\left(1+\frac{b^{2}}{4\vec X^{2}}\right)\frac{\partial\phi(k,\vec X)}{\partial X^{i}}\right)=0.
\end{equation}
By a direct substitution one can show that the scattering states defined as
\begin{align}\label{scatstatesdec0}
\phi_{s}(\vec k,\vec X)&=
\frac{1}{4\pi k}\sum\limits_{l=0}^{\infty}(2l+1)e^{i\left(\frac{\pi l}{2}+\delta_{l,s}(k)\right)}P_{l}\left(\frac{\vec k\vec X}{k|R|}\right)f_{l,s}\left(k,R\right),\\\label{scatstatesdec1}
\phi_{a}(\vec k,\vec X)&=\frac{1}{4\pi k}\sum\limits_{l=0}^{\infty}(2l+1)e^{i\left(\frac{\pi l}{2}+\delta_{l,a}(k)\right)}P_{l}\left(\frac{\vec k\vec X}{k|R|}\right)f_{l,a}\left(k,R\right)
\end{align}
satisfy Eq.~\eqref{eqscalarrad-isotropic-Cart} provided $f_{l,s}\left(k,R\right)$ and $f_{l,a}\left(k,R\right)$ satisfy Eq.~\eqref{eqscalarrad-isotropic}. One can check that at large $|R|$
\begin{align}\label{scatstatesdec1asmp}
\phi_{s}(\vec k,\vec X)&\approx
\frac{1}{\sqrt{2}(2\pi)^{\frac{3}{2}}}\left(e^{i\vec k\vec X}+A_{s}\left(\vec k,\frac{\vec X}{|R|}\right)\frac{e^{ik|R|}}{|R|}\right),\\\label{scatstatesdec2}
\phi_{a}(\vec k,\vec X)&\approx
\frac{\textrm{sign}(R)}{\sqrt{2}(2\pi)^{\frac{3}{2}}}\left(e^{i\vec k\vec X}+A_{a}\left(\vec k,\frac{\vec X}{|R|}\right)\frac{e^{ik|R|}}{|R|}\right),
\end{align}
where the scattering amplitudes $A_{p}(\vec k,\vec n)$ look like
\begin{align}\label{scatampl}
&A_{s}\left(\vec k,\vec n\right)=\frac{1}{2ik}\sum\limits_{l=0}^{\infty}(2l+1)P_{l}\left(\frac{\vec k\vec n}{k}\right)\left(
e^{i2\delta_{l,s}(k)}-1\right),\\
&A_{a}\left(\vec k,\vec n\right)=\frac{1}{2ik}\sum\limits_{l=0}^{\infty}(2l+1)P_{l}\left(\frac{\vec k\vec n}{k}\right)\left(
e^{i2\delta_{l,a}(k)}-1\right).
\end{align}

An important comment is in order. Although the methods of standard scattering theory were used above, one cannot interpret the obtained results within the framework of scattering theory. First, the scalar field squared does not provide the probability density as the absolute value of the wave function squared does in quantum mechanics. Second, there exists the degeneracy of radial solutions with the same $k$ and $l$, leading to the existence of two different scattering states. And third, as mentioned in \S122 of \cite{LL-QM}, ``It is supposed that the incident beam of particles is defined by a wide (to avoid diffraction effects) but finite diaphragm, as happens in actual experiments on scattering''. In contrast to this, since our aim is to build a consistent quantum field theory, we need complete solutions in the whole space.

In this connection, it should be noted that scattering of scalar waves by the Ellis wormhole was examined in \cite{Clement:1982ej}. However, as has already been mentioned, in that paper an extra restriction on the scalar field was imposed which led to discarding radial solutions that are even in $r$. As a result, only one scattering state (which corresponds to scattering state \eqref{scatstatesdec2}) was considered in \cite{Clement:1982ej}.

\section{Physical states}\label{sectnewstates}
Using the scattering states discussed in the previous section, let us consider the following combinations of the scattering states:
\begin{align}\label{phi+def}
&\phi_{+}(\vec k,\vec X)=\frac{1}{\sqrt{2}}\left(\phi_{s}(\vec k,\vec X)+\phi_{a}(\vec k,\vec X)\right),\\\label{phi-def}
&\phi_{-}(\vec k,\vec X)=\frac{1}{\sqrt{2}}\left(\phi_{s}(\vec k,\vec X)-\phi_{a}(\vec k,\vec X)\right).
\end{align}
One can prove (see Appendix) that these new states satisfy the orthogonality conditions
\begin{align}\label{orthscatt2}
&\int\left(1+\frac{b^{2}}{4R^{2}}\right)^{3}\phi_{+}^{*}(\vec k,\vec X)\phi_{-}^{}(\vec k',\vec X)\,d^{3}X=0,\\\label{orthphi+}
&\int\left(1+\frac{b^{2}}{4R^{2}}\right)^{3}\phi_{+}^{*}(\vec k,\vec X)\phi_{+}^{}(\vec k',\vec X)\,d^{3}X=\delta^{(3)}(\vec k-\vec k'),\\\label{orthphi-}
&\int\left(1+\frac{b^{2}}{4R^{2}}\right)^{3}\phi_{-}^{*}(\vec k,\vec X)\phi_{-}^{}(\vec k',\vec X)\,d^{3}X=\delta^{(3)}(\vec k-\vec k'),
\end{align}
where $d^{3}X=d^{3}x$, $R^{2}=\vec x^{2}$ for $R\ge\frac{b}{2}$ and $d^{3}X=d^{3}y$, $R^{2}=\vec y^{2}$ for $R<-\frac{b}{2}$, and the completeness relation
\begin{equation}\label{completepm}
\int\left(\phi_{-}^{*}(\vec k,\vec X)\phi_{-}^{}(\vec k,\vec X')+\phi_{+}^{*}(\vec k,\vec X)\phi_{+}^{}(\vec k,\vec X')\right)d^{3}k=\frac{\delta^{(3)}(\vec X-\vec X')}{\left(1+\frac{b^{2}}{4R^{2}}\right)^{3}}.
\end{equation}
The integration in \eqref{orthscatt2}--\eqref{orthphi-} is performed over both universes. The three-dimensional delta function in \eqref{completepm} should be interpreted as
\begin{equation}
\delta^{(3)}(\vec X-\vec X')=\left[\begin{array}{ll}
\delta^{(3)}(\vec x-\vec x'),\quad\textrm{if}\quad\vec X=\vec x,~\vec X'=\vec x',\\
\delta^{(3)}(\vec y-\vec y'),\quad\textrm{if}\quad\vec X=\vec y,~\vec X'=\vec y',\\
0,\qquad\qquad\quad\,\textrm{if}\quad\vec X=\vec x,~\vec X'=\vec y'\quad\textrm{or}\quad\vec X=\vec y,~\vec X'=\vec x'.
\end{array}
\right.
\end{equation}

Given asymptotics \eqref{scatstatesdec1asmp} and \eqref{scatstatesdec2}, for large positive $R$ the functions $\phi_{+}(\vec k,\vec X)$ and $\phi_{-}(\vec k,\vec X)$ have the form
\begin{align}\label{1overRa0}
&\phi_{+}(\vec k,\vec X)\approx\frac{1}{(2\pi)^{\frac{3}{2}}}\,e^{i\vec k\vec x}
+\frac{1}{2(2\pi)^{\frac{3}{2}}}\left(A_{s}\left(\vec k,\frac{\vec x}{R}\right)+A_{a}\left(\vec k,\frac{\vec x}{R}\right)\right)\frac{e^{ikR}}{R},\\\label{1overRa}
&\phi_{-}(\vec k,\vec X)\approx\frac{1}{2(2\pi)^{\frac{3}{2}}}\left(A_{s}\left(\vec k,\frac{\vec x}{R}\right)-A_{a}\left(\vec k,\frac{\vec x}{R}\right)\right)\frac{e^{ikR}}{R},
\end{align}
whereas for large negative $R$ they have the form
\begin{align}\label{1overRb}
&\phi_{+}(\vec k,\vec X)\approx\frac{1}{2(2\pi)^{\frac{3}{2}}}\left(A_{s}\left(\vec k,\frac{\vec y}{|R|}\right)-A_{a}\left(\vec k,\frac{\vec y}{|R|}\right)\right)\frac{e^{ik|R|}}{|R|},\\\label{1overRb1}
&\phi_{-}(\vec k,\vec X)\approx\frac{1}{(2\pi)^{\frac{3}{2}}}\,e^{i\vec k\vec y}
+\frac{1}{2(2\pi)^{\frac{3}{2}}}\left(A_{s}\left(\vec k,\frac{\vec y}{|R|}\right)+A_{a}\left(\vec k,\frac{\vec y}{|R|}\right)\right)\frac{e^{ik|R|}}{|R|}.
\end{align}
In particular, for $R\to\infty$
\begin{align}\label{phi+infty1}
&\phi_{+}(\vec k,\vec X)\to\frac{1}{(2\pi)^{\frac{3}{2}}}\,e^{i\vec k\vec x},\\
&\phi_{-}(\vec k,\vec X)\to 0,
\end{align}
whereas for $R\to-\infty$
\begin{align}
&\phi_{+}(\vec k,\vec X)\to 0,\\\label{phi+infty2}
&\phi_{-}(\vec k,\vec X)\to\frac{1}{(2\pi)^{\frac{3}{2}}}\,e^{i\vec k\vec y}.
\end{align}
A very important property of \eqref{phi+infty1} and \eqref{phi+infty2} is that these formulas have exactly the same form as the properly normalized plane waves in Minkowski spacetime:
\begin{equation}\label{MinkWF}
\phi(\vec k,\vec x)=\frac{1}{(2\pi)^{\frac{3}{2}}}\,e^{i\vec k\vec x}.
\end{equation}
This property explains the choice of the states $\phi_{+}(\vec k,\vec X)$ and $\phi_{-}(\vec k,\vec X)$. Indeed, since far away from the wormhole the spacetime is almost flat, one may expect that in that area there exists a set of eigenfunctions resembling the complete set of plane waves of Minkowski spacetime. We see that it is exactly the set of eigenfunctions $\phi_{+}(\vec k,\vec X)$ and $\phi_{-}(\vec k,\vec X)$. Owing to such a behavior of eigenfunctions at large $|R|$, the parameter $\vec k$ can be treated as the asymptotic momentum.

The states $\phi_{+}(\vec k,\vec X)$ and $\phi_{-}(\vec k,\vec X)$ possess interesting properties. Let us consider the integrals
\begin{align}\label{normdelta+0}
&\int\left(1+\frac{b^{2}}{4R^{2}}\right)^{3}\phi_{+}^{*}(\vec k,\vec X)\phi_{+}^{}(\vec k,\vec X)\,d^{3}X=\delta^{(3)}(0),\\
&\int\left(1+\frac{b^{2}}{4R^{2}}\right)^{3}\phi_{-}^{*}(\vec k,\vec X)\phi_{-}^{}(\vec k,\vec X)\,d^{3}X=\delta^{(3)}(0),
\end{align}
which follow from the orthogonality conditions \eqref{orthphi+} and \eqref{orthphi-} for $\vec k=\vec k'$. Formally, $\delta^{(3)}(0)$ has an infinite value. However, since $\delta^{(3)}(0)=\frac{1}{(2\pi)^{3}}\int dV=\frac{1}{2\pi^{2}}\int\limits_{0}^{\infty}R^{2}dR$, instead of $\delta^{(3)}(0)$ one can consider the integral $\frac{1}{2\pi^{2}}\int\limits_{0}^{R_{1}}R^{2}dR=\frac{R_{1}^{3}}{6\pi^{2}}$ with large $R_{1}$ as a regularization of three-dimensional delta function. Let us take $0<R_{0}<R_{1}$ such that asymptotics \eqref{1overRa}, \eqref{1overRb}, \eqref{phi+infty1}, \eqref{phi+infty2}, and the relation $\left(1+\frac{b^{2}}{4R^{2}}\right)^{3}\approx 1$ hold with a good accuracy for $|R|>R_{0}$. Then, one gets
\begin{align}\nonumber
&\frac{R_{1}^{3}}{6\pi^{2}}=\int\left(1+\frac{b^{2}}{4R^{2}}\right)^{3}\phi_{+}^{*}(\vec k,\vec X)\phi_{+}^{}(\vec k,\vec X)\,d^{3}X\\\label{normdelta+1}
&\approx\int\limits_{R<R_{0}}\left(1+\frac{b^{2}}{4R^{2}}\right)^{3}\phi_{+}^{*}(\vec k,\vec X)\phi_{+}^{}(\vec k,\vec X)\,d^{3}X+\frac{1}{2\pi^{2}}\int\limits_{R_{0}}^{R_{1}}R^{2}dR
\end{align}
and
\begin{align}\nonumber
&\frac{R_{1}^{3}}{6\pi^{2}}=\int\left(1+\frac{b^{2}}{4R^{2}}\right)^{3}\phi_{-}^{*}(\vec k,\vec X)\phi_{-}^{}(\vec k,\vec X)\,d^{3}X\\\label{normdelta-01}
&\approx\int\limits_{-R_{0}<R}\left(1+\frac{b^{2}}{4R^{2}}\right)^{3}\phi_{-}^{*}(\vec k,\vec X)\phi_{-}^{}(\vec k,\vec X)\,d^{3}X+\frac{1}{2\pi^{2}}\int\limits_{-R_{1}}^{-R_{0}}R^{2}dR.
\end{align}

Now let us consider the standard case of Minkowski spacetime, in which the corresponding eigenfunctions for the scalar field have the form \eqref{MinkWF}. In analogy with \eqref{normdelta+1} and \eqref{normdelta-01}, from the corresponding orthogonality condition one gets
\begin{equation}\label{MinkOrth}
\delta^{(3)}(0)=\int\phi^{*}(\vec k,\vec x)\phi^{}(\vec k,\vec x)\,d^{3}x\to\frac{R_{1}^{3}}{6\pi^{2}}
=\frac{R_{0}^{3}}{6\pi^{2}}+\frac{1}{2\pi^{2}}\int\limits_{R_{0}}^{R_{1}}r^{2}dr.
\end{equation}
For $R_{1}\to\infty$, the first term in the rhs of the second equality in \eqref{MinkOrth} is finite, so in fact it does not provide a relevant contribution to an infinite value of $\delta^{(3)}(0)$. It is the second term in the rhs of the second equality in \eqref{MinkOrth} that provides $\delta^{(3)}(0)$ for $R_{1}\to\infty$. Comparing \eqref{MinkOrth} with \eqref{normdelta+1} and \eqref{normdelta-01}, one expects that the terms
\begin{align}\nonumber
&\int\limits_{R<R_{0}}\left(1+\frac{b^{2}}{4R^{2}}\right)^{3}\phi_{+}^{*}(\vec k,\vec X)\phi_{+}^{}(\vec k,\vec X)\,d^{3}X
=\int\limits_{\frac{b}{2}<R<R_{0}}\left(1+\frac{b^{2}}{4R^{2}}\right)^{3}\phi_{+}^{*}(\vec k,\vec X)\phi_{+}^{}(\vec k,\vec X)\,d^{3}X
\\\label{intfin1}
&+\int\limits_{-R_{0}<R<-\frac{b}{2}}\left(1+\frac{b^{2}}{4R^{2}}\right)^{3}\phi_{+}^{*}(\vec k,\vec X)\phi_{+}^{}(\vec k,\vec X)\,d^{3}X
+\int\limits_{R<-R_{0}}\left(1+\frac{b^{2}}{4R^{2}}\right)^{3}\phi_{+}^{*}(\vec k,\vec X)\phi_{+}^{}(\vec k,\vec X)\,d^{3}X
\end{align}
and
\begin{align}\nonumber
&\int\limits_{-R_{0}<R}\left(1+\frac{b^{2}}{4R^{2}}\right)^{3}\phi_{-}^{*}(\vec k,\vec X)\phi_{-}^{}(\vec k,\vec X)\,d^{3}X
=\int\limits_{-R_{0}<R<-\frac{b}{2}}\left(1+\frac{b^{2}}{4R^{2}}\right)^{3}\phi_{-}^{*}(\vec k,\vec X)\phi_{-}^{}(\vec k,\vec X)\,d^{3}X
\\\label{intfin2}&+
\int\limits_{\frac{b}{2}<R<R_{0}}\left(1+\frac{b^{2}}{4R^{2}}\right)^{3}\phi_{-}^{*}(\vec k,\vec X)\phi_{-}^{}(\vec k,\vec X)\,d^{3}X
+\int\limits_{R_{0}<R}\left(1+\frac{b^{2}}{4R^{2}}\right)^{3}\phi_{-}^{*}(\vec k,\vec X)\phi_{-}^{}(\vec k,\vec X)\,d^{3}X
\end{align}
also do not provide relevant contributions to $\delta^{(3)}(0)$. The second and third terms in the rhs of these formulas describe contributions of the ``opposite'' universes (i.e., universe with $R<-\frac{b}{2}$ for $\phi_{+}(\vec k,\vec X)$ and universe with $R>\frac{b}{2}$ for $\phi_{-}(\vec k,\vec X)$). The first and second terms in the rhs of formulas \eqref{intfin1} and \eqref{intfin2} are finite for $R_{1}\to\infty$, whereas the third terms take the form

\begingroup
\allowdisplaybreaks
\begin{align}\label{oppu1}
&\int\limits_{R<-R_{0}}\left(1+\frac{b^{2}}{4R^{2}}\right)^{3}\phi_{+}^{*}(\vec k,\vec X)\phi_{+}^{}(\vec k,\vec X)\,d^{3}X
\approx
\frac{1}{16\pi^{2}}\int\limits_{-1}^{1}
\left|A_{s}\left(\vec k,\tau\right)-A_{a}\left(\vec k,\tau\right)\right|^{2}d\tau\int\limits_{-R_{1}}^{-R_{0}}dR,\\\label{oppu2}
&\int\limits_{R_{0}<R}\left(1+\frac{b^{2}}{4R^{2}}\right)^{3}\phi_{-}^{*}(\vec k,\vec X)\phi_{-}^{}(\vec k,\vec X)\,d^{3}X
\approx
\frac{1}{16\pi^{2}}\int\limits_{-1}^{1}
\left|A_{s}\left(\vec k,\tau\right)-A_{a}\left(\vec k,\tau\right)\right|^{2}d\tau\int\limits_{R_{0}}^{R_{1}}dR,
\end{align}
\endgroup
where $\tau=\cos\theta$ with $\cos\theta=\frac{\vec k\vec y}{ky}$ in \eqref{oppu1} and $\cos\theta=\frac{\vec k\vec x}{kx}$ in \eqref{oppu2}. Although these integrals $\sim R_{1}$ for large $R_{1}$ (contrary to the corresponding term in \eqref{MinkOrth}), they can be neglected in comparison with the terms in \eqref{normdelta+1} and \eqref{normdelta-01} that are proportional to $R_{1}^{3}$ for large $R_{1}$ (recall that only the first terms in the rhs of \eqref{1overRa0} and \eqref{1overRb1} are retained as the leading asymptotics in the present calculation, whereas the terms $\sim\frac{1}{|R|}$ in \eqref{1overRa0} and \eqref{1overRb1} are neglected --- it is their contribution that is expected to compensate the diverging integrals \eqref{oppu1} and \eqref{oppu2}). Thus, the integrals corresponding to the ``opposite'' universes cannot provide somewhat proportional to $\delta^{(3)}(0)$. In other words, for $\phi_{+}^{}(\vec k,\vec X)$ the three-dimensional delta function $\delta^{(3)}(0)$ is attained in the universe with $R>\frac{b}{2}$, whereas for $\phi_{-}^{}(\vec k,\vec X)$ it is attained in the universe with $R<-\frac{b}{2}$. The reasoning presented above, together with the explicit form of asymptotics \eqref{1overRa0}--\eqref{1overRb1}, suggests that the states $\phi_{+}(\vec k,\vec X)$ live mostly in the universe with $R>\frac{b}{2}$, whereas the states $\phi_{-}(\vec k,\vec X)$ live mostly in the universe with $R<-\frac{b}{2}$; they penetrate into the ``opposite'' universes only to a small extent.

\section{Quantum theory}\label{sectqft}
Now we are ready to build the quantum field theory. The quantum scalar field can be represented as
\begin{align}\nonumber
\phi(t,\vec X)=\int\frac{d^{3}k}{\sqrt{2\sqrt{{\vec k}^{2}+M^{2}}}}\biggl(e^{-i\sqrt{{\vec k}^{2}+M^{2}}\,t}\left(\phi_{-}^{}(\vec k,\vec X)a_{-}^{}(\vec k)+\phi_{+}^{}(\vec k,\vec X)a_{+}^{}(\vec k)\right)&\\\label{operatordec2}+
e^{i\sqrt{{\vec k}^{2}+M^{2}}\,t}\left(\phi_{-}^{*}(\vec k,\vec X)a_{-}^{\dagger}(\vec k)+\phi_{+}^{*}(\vec k,\vec X)a_{+}^{\dagger}(\vec k)\right)\biggr)&,
\end{align}
where the creation and annihilation operators satisfy the standard commutation relations
\begin{align}
&[a_{-}^{}(\vec k),a_{-}^{\dagger}({\vec k}')]=\delta^{(3)}(\vec k-\vec k'),\\
&[a_{+}^{}(\vec k),a_{+}^{\dagger}({\vec k}')]=\delta^{(3)}(\vec k-\vec k'),
\end{align}
all other commutators being equal to zero. Since the functions $\phi_{+}(\vec k,\vec X)$ and $\phi_{-}(\vec k,\vec X)$ form a complete set of eigenfunctions, here they define the one-particle rigged Hilbert space (since the eigenfunctions have an infinite norm, formally one should consider the rigged Hilbert space \cite{rHs} instead of the standard Hilbert space).

In a consistent scalar quantum field theory, the canonical commutation relations
\begin{align}\label{CCR1}
&[\phi(t,\vec X),\pi(t,\vec X')]=i\delta^{(3)}(\vec X-\vec X'),\\\label{CCR2}
&[\phi(t,\vec X),\phi(t,\vec X')]=0,\\\label{CCR3}
&[\pi(t,\vec X),\pi(t,\vec X')]=0
\end{align}
should be exactly satisfied. Here the canonically conjugate momentum is defined in a standard way as
\begin{equation}\label{ccm}
\pi(t,\vec X)=\frac{\partial\mathcal{L}}{\partial\dot\phi(t,\vec X)}=\sqrt{-g(\vec X)}\,g^{00}(\vec X)\dot\phi(t,\vec X)=\left(1+\frac{b^{2}}{4R^{2}}\right)^{3}\dot\phi(t,\vec X).
\end{equation}

First, let us consider commutation relation \eqref{CCR1}. Substituting the field \eqref{operatordec2} into the lhs of \eqref{CCR1}, one gets
\begin{align}\nonumber
[\phi(t,\vec X),\dot\phi(t,\vec X')]&=\frac{1}{2}
\int\Bigl(\phi_{+}^{}(\vec k,\vec X)\phi_{+}^{*}(\vec k,\vec X')+\phi_{+}^{}(\vec k,\vec X')\phi_{+}^{*}(\vec k,\vec X)\\\label{CCR1calc}&+
\phi_{-}^{}(\vec k,\vec X)\phi_{-}^{*}(\vec k,\vec X')+\phi_{-}^{}(\vec k,\vec X')\phi_{-}^{*}(\vec k,\vec X)\Bigr)d^{3}k.
\end{align}
Using the completeness relation \eqref{completepm}, one can easily get \eqref{CCR1} from \eqref{CCR1calc}.

Now let us turn to the commutation relations \eqref{CCR2} and \eqref{CCR3}. Substituting the field \eqref{operatordec2} into \eqref{CCR2} and \eqref{CCR3}, one gets
\begin{align}\nonumber
[\phi(t,\vec X),\phi(t,\vec X')]&=
\frac{1}{2}\int\Bigl(\phi_{+}^{}(\vec k,\vec X)\phi_{+}^{*}(\vec k,\vec X')-\phi_{+}^{}(\vec k,\vec X')\phi_{+}^{*}(\vec k,\vec X)\\\label{CCR2calc0}&+
\phi_{-}^{}(\vec k,\vec X)\phi_{-}^{*}(\vec k,\vec X')-\phi_{-}^{}(\vec k,\vec X')\phi_{-}^{*}(\vec k,\vec X)\Bigr)\frac{d^{3}k}{\sqrt{{\vec k}^{2}+M^{2}}}
\end{align}
and
\begin{align}\nonumber
[\dot\phi(t,\vec X),\dot\phi(t,\vec X')]&=
\frac{1}{2}\int\Bigl(\phi_{+}^{}(\vec k,\vec X)\phi_{+}^{*}(\vec k,\vec X')-\phi_{+}^{}(\vec k,\vec X')\phi_{+}^{*}(\vec k,\vec X)\\\label{CCR3calc0}&+
\phi_{-}^{}(\vec k,\vec X)\phi_{-}^{*}(\vec k,\vec X')-\phi_{-}^{}(\vec k,\vec X')\phi_{-}^{*}(\vec k,\vec X)\Bigr)\sqrt{{\vec k}^{2}+M^{2}}\,d^{3}k.
\end{align}
Using definitions \eqref{phi+def} and \eqref{phi-def}, the latter integrals can be brought to the form
\begin{align}\nonumber
[\phi(t,\vec X),\phi(t,\vec X')]&=
\frac{1}{2}\int\Bigl(\phi_{s}^{}(\vec k,\vec X)\phi_{s}^{*}(\vec k,\vec X')-\phi_{s}^{}(\vec k,\vec X')\phi_{s}^{*}(\vec k,\vec X)\\\label{CCR2calc}&+
\phi_{a}^{}(\vec k,\vec X)\phi_{a}^{*}(\vec k,\vec X')-\phi_{a}^{}(\vec k,\vec X')\phi_{a}^{*}(\vec k,\vec X)\Bigr)\frac{d^{3}k}{\sqrt{{\vec k}^{2}+M^{2}}}
\end{align}
and
\begin{align}\nonumber
[\dot\phi(t,\vec X),\dot\phi(t,\vec X')]&=
\frac{1}{2}\int\Bigl(\phi_{s}^{}(\vec k,\vec X)\phi_{s}^{*}(\vec k,\vec X')-\phi_{s}^{}(\vec k,\vec X')\phi_{s}^{*}(\vec k,\vec X)\\\label{CCR3calc}&+
\phi_{a}^{}(\vec k,\vec X)\phi_{a}^{*}(\vec k,\vec X')-\phi_{a}^{}(\vec k,\vec X')\phi_{a}^{*}(\vec k,\vec X)\Bigr)\sqrt{{\vec k}^{2}+M^{2}}\,d^{3}k.
\end{align}
Next, performing the calculations analogous to those presented in the Appendix (see formulas \eqref{complete4app2} and \eqref{complete4app8} as examples), the integrals in \eqref{CCR2calc} and \eqref{CCR3calc} can be represented as
\begin{align}\nonumber
&\frac{1}{4\pi}\sum\limits_{l=0}^{\infty}(2l+1)
P_{l}(\cos\alpha)\int\limits_{0}^{\infty}\Bigl(
f_{l,s}\left(k,R\right)f_{l,s}\left(k,R'\right)-f_{l,s}\left(k,R'\right)f_{l,s}\left(k,R\right)\\
&+f_{l,a}\left(k,R\right)f_{l,a}\left(k,R'\right)-f_{l,a}\left(k,R'\right)f_{l,a}\left(k,R\right)
\Bigr)\frac{dk}{\sqrt{k^{2}+M^{2}}}=0
\end{align}
and
\begin{align}\nonumber
&\frac{1}{4\pi}\sum\limits_{l=0}^{\infty}(2l+1)
P_{l}(\cos\alpha)\int\limits_{0}^{\infty}\Bigl(
f_{l,s}\left(k,R\right)f_{l,s}\left(k,R'\right)-f_{l,s}\left(k,R'\right)f_{l,s}\left(k,R\right)\\
&+f_{l,a}\left(k,R\right)f_{l,a}\left(k,R'\right)-f_{l,a}\left(k,R'\right)f_{l,a}\left(k,R\right)
\Bigr)\sqrt{k^{2}+M^{2}}\,dk=0.
\end{align}
Thus, we see that canonical commutation relations \eqref{CCR1}--\eqref{CCR3} are exactly satisfied for \eqref{operatordec2}.

Finally, we need to obtain the Hamiltonian of the theory. By definition, the Hamiltonian has the form
\begin{equation}\label{Hamilt}
H=\int\sqrt{-g}\,g^{00}T_{00}d^{3}X.
\end{equation}
Substituting into \eqref{Hamilt} metric \eqref{metric_Sch_isotropic} and the energy-momentum tensor following from action \eqref{scalact}, performing an integration by parts and using the equation of motion for the scalar field, we arrive at
\begin{equation}\label{Hamiltscalar}
H=\frac{1}{2}\int\left(1+\frac{b^{2}}{4R^{2}}\right)^{3}\left(\dot\phi^{2}-\ddot\phi\,\phi\right)d^{3}X.
\end{equation}
Next, substituting representation \eqref{operatordec2} into the latter formula and using the orthogonality conditions \eqref{orthscatt2}--\eqref{orthphi-}, after straightforward calculations we get
\begingroup
\allowdisplaybreaks
\begin{align}\nonumber
H&=\frac{1}{2}\int\sqrt{{\vec k}^{2}+M^{2}}
\left(a_{-}^{\dagger}(\vec k)a_{-}^{}(\vec k)+a_{-}^{}(\vec k)a_{-}^{\dagger}(\vec k)+a_{+}^{\dagger}(\vec k)a_{+}^{}(\vec k)+a_{+}^{}(\vec k)a_{+}^{\dagger}(\vec k)\right)d^{3}k
\\\nonumber
&+\frac{1}{4}\int d^{3}k\int d^{3}k'\left(\frac{({\vec k}^{2}+M^{2})^{\frac{3}{4}}}{({\vec {k'}}^{2}+M^{2})^{\frac{1}{4}}}
-({\vec k}^{2}+M^{2})^{\frac{1}{4}}({\vec {k'}}^{2}+M^{2})^{\frac{1}{4}}\right)
\Biggl(e^{-i\left(\sqrt{{\vec k}^{2}+M^{2}}+\sqrt{{\vec {k'}}^{2}+M^{2}}\right)t}\\\nonumber&\times\int\left(1+\frac{b^{2}}{4R^{2}}\right)^{3}
\left(\phi_{+}^{}(\vec k,\vec X)a_{+}^{}(\vec k)+\phi_{-}^{}(\vec k,\vec X)a_{-}^{}(\vec k)\right)\\\label{Hamaux1}&\times\left(\phi_{+}^{}(\vec k',\vec X)a_{+}^{}(\vec k')+\phi_{-}^{}(\vec k',\vec X)a_{-}^{}(\vec k')\right)\,d^{3}X+\textrm{H.c.}\Biggr).
\end{align}
\endgroup
Taking into account definitions \eqref{phi+def} and \eqref{phi-def}, integration over $\vec X$ in \eqref{Hamaux1} is reduced to calculation of the integrals
\begin{equation}\label{orthcondaux2-0}
\int\left(1+\frac{b^{2}}{4R^{2}}\right)^{3}
\phi_{p}^{}(\vec k,\vec X)\phi_{p'}^{}(\vec k',\vec X)\,d^{3}X,
\end{equation}
where $p,p'=s,a$. Again, performing calculations analogous to those presented in the Appendix (see formulas \eqref{orthscatt2app} and \eqref{orthscattfinapp} as examples), it is not difficult to show that
\begin{equation}\label{orthcondaux2}
\int\left(1+\frac{b^{2}}{4R^{2}}\right)^{3}
\phi_{p}^{}(\vec k,\vec X)\phi_{p'}^{}(\vec k',\vec X)\,d^{3}X=\frac{1}{4\pi k^{2}}\sum\limits_{l=0}^{\infty}(2l+1)e^{i(\pi l+2\delta_{l,p}(k))}P_{l}(\cos\alpha)\delta_{pp'}\delta(k-k'),
\end{equation}
where $\alpha$ is the angle between the vectors $\vec k$ and $\vec k'$, $k=\sqrt{{\vec k}^{2}}$, and $k'=\sqrt{{\vec {k'}}^{2}}$. Since formula \eqref{orthcondaux2} contains $\delta(k-k')$, we get
\begin{equation}
\left(\frac{({\vec k}^{2}+M^{2})^{\frac{3}{4}}}{({\vec {k'}}^{2}+M^{2})^{\frac{1}{4}}}
-({\vec k}^{2}+M^{2})^{\frac{1}{4}}({\vec {k'}}^{2}+M^{2})^{\frac{1}{4}}\right)\delta(k-k')=\left(\sqrt{k^{2}+M^{2}}-\sqrt{k^{2}+M^{2}}\right)\delta(k-k')=0,
\end{equation}
which brings \eqref{Hamaux1} to the form
\begin{equation}\label{Hamaux1b}
H=\frac{1}{2}\int\sqrt{{\vec k}^{2}+M^{2}}
\left(a_{-}^{\dagger}(\vec k)a_{-}^{}(\vec k)+a_{-}^{}(\vec k)a_{-}^{\dagger}(\vec k)+a_{+}^{\dagger}(\vec k)a_{+}^{}(\vec k)+a_{+}^{}(\vec k)a_{+}^{\dagger}(\vec k)\right)d^{3}k.
\end{equation}
The latter formula can be brought to the form
\begin{equation}\label{Hamiltresult}
H=\int\sqrt{{\vec k}^{2}+M^{2}}
\left(a_{-}^{\dagger}(\vec k)a_{-}^{}(\vec k)+a_{+}^{\dagger}(\vec k)a_{+}^{}(\vec k)\right)d^{3}k,
\end{equation}
where infinite c-number terms are dropped. One can see that this Hamiltonian differs from the standard Hamiltonian of a real massive scalar field in Minkowski spacetime
\begin{equation}\label{HamiltscfM}
H=\int\sqrt{{\vec k}^{2}+M^{2}}\,a^{\dagger}(\vec k)a^{}(\vec k)\,d^{3}k
\end{equation}
by the degeneracy of states with the same vector $\vec k$ in \eqref{Hamiltresult}.

\section{Conclusion}\label{sectconclusion}
In the present paper, quantization of a real massive scalar field in the background of a traversable wormhole connecting two different universes is examined. As an example, the simplest Ellis wormhole is considered. The equation of motion for the scalar field is examined in detail, and a complete set of eigenfunctions is presented. An important property of these eigenfunctions is that far away from the wormhole they look just as properly normalized plane waves. Moreover, the complete set of eigenfunctions is split into two sets (consisting of the states $\phi_{+}(\vec k,\vec X)$ and $\phi_{-}(\vec k,\vec X)$ respectively), each of which lives mostly in its own universe and penetrates into the ``opposite'' universe only to a small extent. Decomposition of the quantum scalar field into these states is performed, validity of the corresponding canonical quantization relations is verified, and the Hamiltonian of the resulting theory is obtained. It is shown that this Hamiltonian does not contain any pathologies. However, a remarkable feature of the resulting theory is the doubling of quantum states parametrized by the same asymptotic momentum $\vec k$ (these are the states $\phi_{+}(\vec k,\vec X)$ and $\phi_{-}(\vec k,\vec X)$). This degeneracy is a consequence of $R^{2}\times S^{2}$ topology of the spacetime under consideration; it is expected for other types of wormholes connecting different universes too. Exactly the same degeneracy exists in the Schwarzschild spacetime, which has the same topological structure \cite{Egorov:2022hgg,Smolyakov:2023pml}.

In this connection, it is worth discussing compatibility of some of the earlier results on quantum field theory in the wormhole background (in particular, calculations of the two-point correlation function and the expectation value of the energy-momentum tensor) with the degeneracy described above.\footnote{I am grateful to the anonymous referee for the suggestion to address this point.} Calculations of renormalized $\langle\phi^{2}\rangle$ and $\langle T_{\mu\nu}\rangle$ for various wormholes, which are based on the method developed in \cite{Anderson:1990jh,Anderson:1993if,Anderson:1994hg} for spherically symmetric spacetimes and applied to black holes in these papers, can be found in \cite{Hochberg:1996ee,Taylor:1996yu,Popov:2000id,Carlson:2010yw}. It turns out that the results obtained in \cite{Hochberg:1996ee,Taylor:1996yu,Popov:2000id,Carlson:2010yw} are not affected by the degeneracy. Indeed, the degeneracy in \eqref{Hamiltresult} is a direct consequence of the existence of two linearly independent radial solutions for given $k$ and $l$, which are considered as physical ones (in the present study, these solutions are chosen to be symmetric and antisymmetric in $r$; see \eqref{radsols} and \eqref{radsola}). The point is that the method developed in \cite{Anderson:1990jh,Anderson:1993if,Anderson:1994hg} relies on the use of two linearly independent solutions of the corresponding radial equation of motion for given $k$ (in our notations) and $l$, even if they diverge for some $r$ \cite{Anderson:1990jh} and thus cannot be considered as physical solutions (recall that in the case of Minkowski spacetime for given $k$ and $l$ formally there exist two linearly independent solutions of the radial equation of motion, but only one of them is finite for all $r$ and can be considered as a physical one \cite{LL-QM}, whereas in the case of wormholes both solutions are finite for all $r$). Thus, the doubling of physical radial solutions for wormholes connecting two different universes in comparison with the case of Minkowski spacetime is taken into account in \cite{Hochberg:1996ee,Taylor:1996yu,Popov:2000id,Carlson:2010yw} due to the specificity of the method itself. In particular, $\langle\phi^{2}\rangle$ for the wormhole with metric \eqref{metric_WH} (including the case of minimal coupling of the scalar field to gravity, which is considered in the present paper) was calculated analytically in \cite{Popov:2000id}, so if we somehow recalculated $\langle\phi^{2}\rangle$ using the quantum field \eqref{operatordec2} instead of the field expanded in spherical harmonics, we would get the result of \cite{Popov:2000id}.  Thus, although the degeneracy in terms of the states $\phi_{+}(\vec k,\vec X)$ and $\phi_{-}(\vec k,\vec X)$, which are familiar to a distant observer because far away from the wormhole they look like standard plane waves, was not identified explicitly in \cite{Hochberg:1996ee,Taylor:1996yu,Popov:2000id,Carlson:2010yw}; it was taken into account automatically.

Since the states $\phi_{+}(\vec k,\vec X)$ and $\phi_{-}(\vec k,\vec X)$ live in different universes, there arises a question of passing of particles through such a wormhole (i.e., from one universe to another). In principle, the problem can be formulated even at the classical level: which part of a wave packet composed, say, only from the modes $\phi_{+}(\vec k,\vec X)$ can pass through the wormhole into a different universe taking into account that the modes living in the destination universe are $\phi_{-}(\vec k,\vec X)$, not $\phi_{+}(\vec k,\vec X)$.\footnote{The problem looks considerably different if there exist nonlinear interactions, because in such a case the modes $\phi_{-}(\vec k,\vec X)$ can be created from the modes $\phi_{+}(\vec k,\vec X)$ due to nonlinear interactions. Moreover, there may exist objects like solitons that do not spread out during their motion.} This problem calls for a further investigation.

\subsection*{Acknowledgments}
The author is grateful to the anonymous referee of paper \cite{Egorov:2022hgg} whose comment led to the present paper, and to V.O.~Egorov and I.P.~Volobuev for useful comments. This study was conducted within the scientific program of the National Center for Physics and Mathematics, Section No.~5 ``Particle Physics and Cosmology'', stage 2023-2025.

\section*{Appendix: Orthogonality conditions and completeness relation for $\phi_{\pm}(\vec k,\vec X)$}
In order to prove the validity of orthogonality conditions \eqref{orthscatt2}--\eqref{orthphi-}, let us adopt the method used in \S134 of \cite{LL-QM} in the case of standard scattering states. Substituting \eqref{phi+def} and \eqref{phi-def} into the lhs of \eqref{orthscatt2}--\eqref{orthphi-}, one gets
\begingroup
\allowdisplaybreaks
\begin{align}\nonumber
\frac{1}{2}\int d^{3}X\left(1+\frac{b^{2}}{4R^{2}}\right)^{3}&\Bigl(\phi_{s}^{*}(\vec k,\vec X)\phi_{s}^{}(\vec k',\vec X)-\phi_{a}^{*}(\vec k,\vec X)\phi_{a}^{}(\vec k',\vec X)\\\label{orthscatt2app1}
&-\phi_{s}^{*}(\vec k,\vec X)\phi_{a}^{}(\vec k',\vec X)+\phi_{a}^{*}(\vec k,\vec X)\phi_{s}^{}(\vec k',\vec X)\Bigr),\\\nonumber
\frac{1}{2}\int d^{3}X\left(1+\frac{b^{2}}{4R^{2}}\right)^{3}&\Bigl(\phi_{s}^{*}(\vec k,\vec X)\phi_{s}^{}(\vec k',\vec X)+\phi_{a}^{*}(\vec k,\vec X)\phi_{a}^{}(\vec k',\vec X)\\\label{orthscatt2app2}
&+\phi_{s}^{*}(\vec k,\vec X)\phi_{a}^{}(\vec k',\vec X)+\phi_{a}^{*}(\vec k,\vec X)\phi_{s}^{}(\vec k',\vec X)\Bigr),\\\nonumber
\frac{1}{2}\int d^{3}X\left(1+\frac{b^{2}}{4R^{2}}\right)^{3}&\Bigl(\phi_{s}^{*}(\vec k,\vec X)\phi_{s}^{}(\vec k',\vec X)+\phi_{a}^{*}(\vec k,\vec X)\phi_{a}^{}(\vec k',\vec X)\\\label{orthscatt2app3}
&-\phi_{s}^{*}(\vec k,\vec X)\phi_{a}^{}(\vec k',\vec X)-\phi_{a}^{*}(\vec k,\vec X)\phi_{s}^{}(\vec k',\vec X)\Bigr).
\end{align}
\endgroup
Thus, calculation of the integrals in \eqref{orthscatt2app1}--\eqref{orthscatt2app3} is reduced to calculation of the integrals
\begin{equation}\label{intortappA}
\int d^{3}X\left(1+\frac{b^{2}}{4R^{2}}\right)^{3}\phi_{p}^{*}(\vec k,\vec X)\phi_{p'}^{}(\vec k',\vec X),
\end{equation}
where $p,p'=s,a$.

Let us take $\phi_{p}(\vec k,\vec X)$ defined by \eqref{scatstatesdec0} or \eqref{scatstatesdec1}. Let $\beta$ be the angle between the vectors $\vec k$ and $\vec X$, $\beta'$ be the angle between the vectors $\vec k'$ and $\vec X$, $\alpha$ be the angle between the vectors $\vec k$ and $\vec k'$, and $\tilde\varphi$ be the angle between the planes $(\vec X,\vec k)$ and $(\vec k,\vec k')$. It is well known that in such a case the relation
\begin{equation}
\cos\beta'=\cos\beta\cos\alpha+\sin\beta\sin\alpha\cos\tilde\varphi
\end{equation}
and the addition theorem for Legendre polynomials
\begin{equation}\label{addtheorem}
P_{l'}(\cos\beta')=P_{l'}(\cos\beta)P_{l'}(\cos\alpha)
+2\sum\limits_{m=1}^{l'}\frac{(l'-m)!}{(l'+m)!}P_{l'}^{m}(\cos\beta)P_{l'}^{m}(\cos\alpha)\cos(m\tilde\varphi)
\end{equation}
hold \cite{Korn-Korn}. Since the integrations with respect to $\beta$ and $\tilde\varphi$ go over the total solid angle of the vector $\vec X$, the integral \eqref{intortappA} can be rewritten as
\begin{align}\nonumber
&\frac{1}{16\pi^{2}kk'}\sum\limits_{l=0}^{\infty}
\sum\limits_{l'=0}^{\infty}(2l+1)(2l'+1)e^{i\left(\frac{\pi(l-l')}{2}+\delta_{l,p}(k)-\delta_{l',p'}(k')\right)}
\\\label{orthscatt2app}&\times
\int\limits_{0}^{2\pi}d\tilde\varphi\int\limits_{0}^{\pi}\sin\beta\,d\beta P_{l}(\cos\beta)P_{l'}(\cos\beta')\int\limits_{\hat L}dR\,R^{2}\left(1+\frac{b^{2}}{4R^{2}}\right)^{3}
f_{l,p}\left(k,R\right)f_{l',p'}\left(k',R\right),
\end{align}
where $\hat L=\left(-\infty,-\frac{b}{2}\right)\cup\left[\frac{b}{2},\infty\right)$. Substituting \eqref{addtheorem} into \eqref{orthscatt2app}, we get
\begin{align}\nonumber
&\sum\limits_{l=0}^{\infty}\frac{2l+1}{4\pi kk'}P_{l}(\cos\alpha)e^{i\left(\delta_{l,p}(k)-\delta_{l,p'}(k')\right)}
\int\limits_{\hat L}dR\frac{\left(4R^{2}+b^{2}\right)^{3}}{64R^{4}}
f_{l,p}\left(k,R\right)f_{l,p'}\left(k',R\right)\\\label{orthscattfinapp}
&=\delta_{pp'}\delta(k-k')\sum\limits_{l=0}^{\infty}\frac{2l+1}{4\pi k^{2}}P_{l}(\cos\alpha)=\delta_{pp'}\frac{1}{\pi k^{2}}\,\delta(k-k')\delta(1-\cos\alpha),
\end{align}
where orthogonality conditions \eqref{normqК1}--\eqref{normqК3} and the relation \cite{LL-QM}
\begin{equation}\label{Pdelta}
\frac{1}{4}\sum\limits_{l=0}^{\infty}(2l+1)P_{l}(\cos\alpha)=\delta(1-\cos\alpha)
\end{equation}
were used.

The delta functions in the last term of \eqref{orthscattfinapp} select exactly $\vec k=\vec k'$. One can prove that the rhs of \eqref{orthscattfinapp} is just $\delta^{(3)}(\vec k-\vec k')$. To do it, let us consider a three-dimensional integral over $\vec k$ such that the ``$z$ axis'' of the corresponding spherical coordinate system coincides with the direction of vector $\vec k'$ \cite{LL-QM}:
\begin{align}\nonumber
&\int d^{3}k\left(\frac{1}{\pi k^{2}}\,\delta(k-k')\delta(1-\cos\alpha)\right)
\\\label{orthscatt2app5}
&=2\pi\int\limits_{0}^{\infty}k^{2}dk\int\limits_{0}^{\pi}\sin\alpha\,d\alpha\left(\frac{1}{\pi k^{2}}\,\delta(k-k')\delta(1-\cos\alpha)\right)=2\int\limits_{-1}^{1}\delta(1-y)dy=1.
\end{align}
Here the prescription
\begin{equation}\label{prescr}
\int_{-1}^{1}\delta(1-y)dy=\frac{1}{2}
\end{equation}
is used. Substituting \eqref{orthscattfinapp} into \eqref{orthscatt2app1}--\eqref{orthscatt2app3} leads to \eqref{orthscatt2}--\eqref{orthphi-}.

Now let us turn to the completeness relation. The lhs of formula \eqref{completepm} can be rewritten as
\begin{align}\nonumber
&\int\left(\phi_{-}^{*}(\vec k,\vec X)\phi_{-}^{}(\vec k,\vec X')+\phi_{+}^{*}(\vec k,\vec X)\phi_{+}^{}(\vec k,\vec X')\right)d^{3}k\\\label{compapp}
&=\int\left(\phi_{s}^{*}(\vec k,\vec X)\phi_{s}^{}(\vec k,\vec X')+\phi_{a}^{*}(\vec k,\vec X)\phi_{a}^{}(\vec k,\vec X')\right)d^{3}k.
\end{align}
Let $\beta$ be the angle between the vectors $\vec X$ and $\vec k$, $\beta'$ be the angle between the vectors $\vec X'$ and $\vec k$, $\alpha$ be the angle between the vectors $\vec X$ and $\vec X'$, and $\tilde\varphi$ be the angle between the planes $(\vec k,\vec X)$ and $(\vec X,\vec X')$. Then, in analogy with formula \eqref{orthscatt2app}, the rhs of \eqref{compapp} can be rewritten as
\begin{align}\nonumber
&\frac{1}{16\pi^{2}}\sum\limits_{l=0}^{\infty}\sum\limits_{l'=0}^{\infty}(2l+1)(2l'+1)e^{i\frac{\pi(l-l')}{2}}
\int\limits_{0}^{2\pi}d\tilde\varphi\int\limits_{0}^{\pi}\sin\beta\,d\beta P_{l}(\cos\beta)P_{l'}(\cos\beta')\\\label{complete4app2}&\times\left(\int\limits_{0}^{\infty}dk\,e^{i\left(\delta_{l,s}(k)-\delta_{l',s}(k)\right)}
f_{l,s}\left(k,R\right)f_{l',s}\left(k,R'\right)+
\int\limits_{0}^{\infty}dk\,e^{i\left(\delta_{l,a}(k)-\delta_{l',a}(k)\right)}
f_{l,a}\left(k,R\right)f_{l',a}\left(k,R'\right)\right).
\end{align}
Using \eqref{addtheorem}, \eqref{Pdelta}, and \eqref{comp2}, formula \eqref{complete4app2} can be brought to the form
\begin{align}\nonumber
&\frac{1}{4\pi}\sum\limits_{l=0}^{\infty}(2l+1)P_{l}(\cos\alpha)\int\limits_{0}^{\infty}dk\Bigl(f_{l,s}\left(k,R\right)f_{l,s}\left(k,R'\right)
+f_{l,a}\left(k,R\right)f_{l,a}\left(k,R'\right)\Bigr)\\\label{complete4app8}
&=\frac{1}{4\pi}\sum\limits_{l=0}^{\infty}(2l+1)P_{l}(\cos\alpha)\frac{64R^{4}}{\left(4R^{2}+b^{2}\right)^{3}}\,\delta(R-R')
=\frac{1}{\pi}\,\delta(1-\cos\alpha)\frac{64R^{4}}{\left(4R^{2}+b^{2}\right)^{3}}\,\delta(R-R').
\end{align}
As in \eqref{orthscattfinapp}, the delta functions in the last term of \eqref{complete4app8} select exactly $\vec X=\vec X'$ (more precisely, $\vec x=\vec x'$ if $R'>\frac{b}{2}$ or $\vec y=\vec y'$ if $R'<-\frac{b}{2}$). To prove that the rhs of \eqref{complete4app8} provides $\delta^{(3)}(\vec X-\vec X')$, let us consider a three-dimensional integral over $\vec X$ such that the ``$z$ axis'' of the corresponding spherical coordinate system coincides with the direction of vector $\vec X'$:
\begin{align}\nonumber
&\int d^{3}X\left(1+\frac{b^{2}}{4R^{2}}\right)^{3}\left(\frac{1}{\pi}\,\delta(1-\cos\alpha)\frac{64R^{4}}{\left(4R^{2}+b^{2}\right)^{3}}\,\delta(R-R')\right)\\\nonumber
&=2\pi\int\limits_{\hat L}dR\,R^{2}\left(1+\frac{b^{2}}{4R^{2}}\right)^{3}\int\limits_{0}^{\pi}d\alpha\sin\alpha
\left(\frac{1}{\pi}\,\delta(1-\cos\alpha)\frac{64R^{4}}{\left(4R^{2}+b^{2}\right)^{3}}\,\delta(R-R')\right)\\\label{complete4app10}
&=2\int\limits_{-1}^{1}\delta(1-y)dy=1.
\end{align}
Here the prescription \eqref{prescr} is also used, and $\hat L$ is defined just after formula \eqref{orthscatt2app}. Thus, \eqref{complete4app8} with \eqref{complete4app10} is just \eqref{completepm}.


\begin{thebibliography}{99}
\bibitem{BD}
N.D.~Birrell, P.C.W.~Davies, {\em ``Quantum fields in curved space''}, Cambridge University Press, Cambridge (1984).

\bibitem{Boulware:1974dm}
D.G.~Boulware, {\em ``Quantum field theory in Schwarzschild and Rindler spaces''},
Phys. Rev. D \textbf{11} (1975) 1404.

\bibitem{HH}
J.B.~Hartle, S.W.~Hawking, {\em ``Path-integral derivation of black-hole radiance''}, Phys. Rev. D \textbf{13} (1976) 2188.

\bibitem{Ellis:1973yv}
H.G.~Ellis, {\em ``Ether flow through a drainhole: A particle model in general relativity''}, J. Math. Phys. \textbf{14} (1973) 104.

\bibitem{Bronnikov:1973fh}
K.A.~Bronnikov, {\em ``Scalar-tensor theory and scalar charge''}, Acta Phys. Polon. B \textbf{4} (1973) 251.

\bibitem{Morris:1988cz}
M.S.~Morris, K.S.~Thorne, {\em ``Wormholes in spacetime and their use for interstellar travel: A tool for teaching general relativity''},
Am. J. Phys. \textbf{56} (1988) 395.

\bibitem{Egorov:2022hgg}
V.~Egorov, M.~Smolyakov, I.~Volobuev, {\em ``Doubling of physical states in the quantum scalar field theory for a remote observer in the Schwarzschild spacetime''}, Phys. Rev. D \textbf{107} (2023) 025001 [arXiv:2209.02067 [gr-qc]].

\bibitem{Smolyakov:2023pml}
M.N.~Smolyakov, {\em ``Asymptotic behavior of solutions and spectrum of states in the quantum scalar field theory in the Schwarzschild spacetime''},
Phys. Rev. D \textbf{108} (2023) 105006 [arXiv:2309.06249 [gr-qc]].

\bibitem{Korn-Korn}
G.A.~Korn, T.M.~Korn, {\em ``Mathematical handbook for scientists and engineers''}, McGraw-Hill, New York (1968).

\bibitem{Clement:1982ej}
G.~Clement, {\em ``Scattering of Klein-Gordon and Maxwell waves by an Ellis geometry''}, Int. J. Theor. Phys. \textbf{23} (1984) 335.

\bibitem{Kar:1994ty}
S.~Kar, D.~Sahdev, B.~Bhawal, {\em ``Scalar waves in a wormhole geometry''}, Phys. Rev. D \textbf{49} (1994) 853.

\bibitem{LL-QM}
L.D.~Landau, E.M.~Lifshitz, {\em ``Quantum mechanics. Non-relativistic theory''}, 2nd ed., Pergamon press, New York (1965).

\bibitem{rHs}
J.-P.~Antoine, R.C.~Bishop, A.~Bohm, S.~Wickramasekara, {\em ``Rigged Hilbert spaces in quantum physics''}, In: D.~Greenberger, K.~Hentschel, F.~Weinert (eds), {\em ``Compendium of quantum physics''}, Springer, Berlin, Heidelberg (2009).

\bibitem{Anderson:1990jh}
P.R.~Anderson, {\em ``A method to compute $\langle\phi^{2}\rangle$ in asymptotically flat, static, spherically symmetric spacetimes''},
Phys. Rev. D \textbf{41} (1990) 1152.

\bibitem{Anderson:1993if}
P.R.~Anderson, W.A.~Hiscock, D.A.~Samuel, {\em ``Stress-energy tensor of quantized scalar fields in static black hole spacetimes''},
Phys. Rev. Lett. \textbf{70} (1993) 1739.

\bibitem{Anderson:1994hg}
P.R.~Anderson, W.A.~Hiscock, D.A.~Samuel, {\em ``Stress-energy tensor of quantized scalar fields in static spherically symmetric spacetimes''},
Phys. Rev. D \textbf{51} (1995) 4337.

\bibitem{Hochberg:1996ee}
D.~Hochberg, A.~Popov, S.V.~Sushkov, {\em ``Selfconsistent wormhole solutions of semiclassical gravity''},
Phys. Rev. Lett. \textbf{78} (1997) 2050 [arXiv:gr-qc/9701064 [gr-qc]].

\bibitem{Taylor:1996yu}
B.E.~Taylor, W.A.~Hiscock, P.R.~Anderson, {\em ``Stress-energy of a quantized scalar field in static wormhole spacetimes''},
Phys. Rev. D \textbf{55} (1997) 6116 [arXiv:gr-qc/9608036 [gr-qc]].

\bibitem{Popov:2000id}
A.A.~Popov, S.V.~Sushkov, {\em ``Vacuum polarization of a scalar field in wormhole spacetimes''},
Phys. Rev. D \textbf{63} (2001) 044017 [arXiv:gr-qc/0009028 [gr-qc]].

\bibitem{Carlson:2010yw}
E.D.~Carlson, P.R.~Anderson, A.~Fabbri, S.~Fagnocchi, W.H.~Hirsch, S.~Klyap, {\em ``Semiclassical gravity in the far field limit of stars, black holes, and wormholes''}, Phys. Rev. D \textbf{82} (2010) 124070 [arXiv:1008.1433 [gr-qc]].
\end{thebibliography}
\end{document}